\documentclass[twocolumn,aps,prc,superscriptaddress,showpacs,floatfix,longbibliography,nofootinbib]{revtex4-1}
\usepackage{url}
\usepackage{cancel}
\usepackage[colorlinks,linkcolor=blue,citecolor=blue,filecolor=black,urlcolor=blue]{hyperref}
\usepackage{epsfig,graphics}
\usepackage{graphicx}% Include figure files
\usepackage{dcolumn}% Align table columns on decimal point
\usepackage{bm}% bold math
\usepackage[usenames]{color}
\usepackage{amssymb}
\usepackage{amsmath}
\usepackage{multirow}
\usepackage{float}
\usepackage{harpoon}
\usepackage{MnSymbol}
\usepackage{appendix}
\usepackage{color}
\usepackage{hyperref}
\usepackage{cleveref}

\begin{document}

\title{Bayesian inference of neutron-skin thickness and neutron-star observables based on effective nuclear interactions}
\author{Jia Zhou}
\affiliation{Shanghai Institute of Applied Physics, Chinese Academy of Sciences, Shanghai 201800, China}
\affiliation{University of Chinese Academy of Sciences, Beijing 100049, China}
\author{Jun Xu}\email[Correspond to\ ]{junxu@tongji.edu.cn}
\affiliation{School of Physics Science and Engineering, Tongji University, Shanghai 200092, China}
%\affiliation{Shanghai Advanced Research Institute, Chinese Academy of Sciences, Shanghai 201210, China}
\affiliation{Shanghai Institute of Applied Physics, Chinese Academy of Sciences, Shanghai 201800, China}
\begin{abstract}
We have obtained the constraints on the density dependence of the symmetry energy from neutron-skin thickness data by parity-violating electron scatterings and neutron-star observables using a Bayesian approach, based on the standard Skyrme-Hartree-Fock (SHF) model and its extension as well as the relativistic mean-field (RMF) model. While the neutron-skin thickness data (neutron-star observables) mostly constrain the symmetry energy at subsaturation (suprasaturation) densities, they may more or less constrain the behavior of the symmetry energy at suprasaturation (subsaturation) densities, depending on the energy-density functional form. Besides showing the final posterior density dependence of the symmetry energy, we also compare the slope parameters of the symmetry energy at 0.10 fm$^{-3}$ as well as the values of the symmetry energy at twice saturation density from three effective nuclear interactions. The present work serves as a comparison study based on relativistic and non-relativistic energy-density functionals, for constraining the nuclear symmetry energy from low to high densities using a Bayesian approach.

\end{abstract}
\maketitle

\section{Introduction}
\label{sec:intro}

The nuclear symmetry energy $E_{sym}(\rho)$ is one of the most uncertain part of the nuclear matter equation of state (EOS), and great efforts have been devoted to extract its density dependence in the past twenty years~\cite{Steiner:2004fi,Lattimer:2006xb,Li:2008gp}. While the nuclear symmetry energy may affect properties of various nuclear systems from finite nuclei to neutron stars~\cite{Steiner:2004fi,Lattimer:2006xb}, different observables are sensitive to the $E_{sym}(\rho)$ at different density regions~\cite{Lynch:2021xkq}. For example, the neutron-skin thickness of a nucleus is most sensitive to the slope parameter of the $E_{sym}(\rho)$ around $\rho=2\rho_0/3$~\cite{Zhang:2013wna,Xu:2020fdc}, with $\rho_0$ being the saturation density, while the radius of a neutron star is most sensitive to the $E_{sym}(\rho)$ around and above $\rho_0$~\cite{Lattimer:2006xb,PhysRevC.103.065804,PhysRevD.106.083010}. Combining the data of neutron-skin thickness and neutron stars may help to constrain the nuclear matter EOS, particularly the $E_{sym}(\rho)$, from low to high densities~\cite{PhysRevC.101.034303,PhysRevC.103.034330,PhysRevC.108.025809,PhysRevC.107.015803,PhysRevC.107.045802}.

The recent PREX and CREX experiments have provided the data of the neutron-skin thickness for $^{208}$Pb~\cite{PREX:2021umo} and $^{48}$Ca~\cite{CREX:2022kgg}, with the former (latter) favoring a large (small) slope parameter $L$ of the $E_{sym}(\rho)$. While the measurement through the parity-violating electron-nucleus scatterings is less model-dependent, the large error bars of the corresponding experimental data may hamper us from putting a strong constraint on the $L$, and it is of interest to see how $E_{sym}(\rho)$ is constrained from both PREX and CREX data. Besides the neutron-skin thickness from light to heavy nuclei, the emergence of recent neutron-star observables, especially neutron-star radii, provides good opportunities to constrain the $E_{sym}(\rho)$ at suprasaturation densities. Recently, the GW170817 event~\cite{LIGOScientific:2018cki} analyzed by the LIGO Scientific Collaboration and Virgo Collaboration as well as the PSR J0030+0451~\cite{Riley:2019yda} and PSR J0740+6620~\cite{Riley_2021} measured by NICER have provided high-quality data for both neutron-star radii and masses, putting constraints on $E_{sym}(\rho)$ at high densities characterized by not only the slope parameter $L$ but also higher-order EOS parameters (see, e.g., Refs.~\cite{Li:2021thg,Lattimer:2021emm,Huth:2021bsp}).

To take good use of many data sets in various systems from finite nuclei to neutron stars including those mentioned above, the Bayesian analysis serves as a good tool to give quantitative constraints on model parameters. On the other hand, the results of such analysis generally depend on the particular theoretical model employed in the study, which uses model parameters as input and provides results of observbles as output. In the present study, we employ non-relativistic and relativistic effective nuclear interactions, in order to check with the model dependence of the constraints on the $E_{sym}(\rho)$. For the non-relativistic effective nuclear interaction, we use the standard SHF model as well as its extension, i.e., the Korea-IBS-Daegu-SKKU (KIDS) model. For the relativistic effective nuclear interaction, we use the relativistic mean-field (RMF) model with $\sigma$, $\omega$, and $\rho$ mesons. The advantages of using these energy-density functionals (EDFs) is that one can express inversely model coefficients in terms of macroscopic physics quantities~\cite{Chen:2010qx,PhysRevC.105.044305,PhysRevC.90.044305}. In that case, one can then set these macroscopic physics quantities as model parameters, so that the sampling in the parameter space becomes more efficient in the Bayesian analysis. The present study could be considered as one of the applications of the machine learning in nuclear physics (see, e.g., Refs.~\cite{Cao23,Sha22,He23a,Gao23,Li23,He23b,Ma23}).

The rest part of the paper is organized as follows. Section~\ref{sec:theory} provides briefly the theoretical framework, including an introduction of the standard SHF model as well as its extension and the RMF model, the calculation method of nucleon density distributions in finite nuclei as well as the mass-radius relation of neutron stars, and the Bayesian analysis method. Section~\ref{sec:results} shows the resulting constraints on the parameters of the $E_{sym}(\rho)$ as well as its density dependence from different observables based on three effective nuclear interactions using the Bayesian approach. We conclude and outlook in Sec.~\ref{sec:summary}.

\section{Theoretical framework}
\label{sec:theory}

In the standard SHF model, the effective interaction between two nucleons at the positions $\vec{r}_1$ and $\vec{r}_2$ is expressed as
\begin{eqnarray}\label{SHFI}
v^{SHF}(\vec{r}_1,\vec{r}_2) &=& t_0(1+x_0P_\sigma)\delta(\vec{r}) \notag \\
&+& \frac{1}{2} t_1(1+x_1P_\sigma)[{\vec{k}'^2}\delta(\vec{r})+\delta(\vec{r})\vec{k}^2] \notag\\
&+&t_2(1+x_2P_\sigma)\vec{k}' \cdot \delta(\vec{r})\vec{k} \notag\\
&+&\frac{1}{6}t_3(1+x_3P_\sigma)\rho^\alpha(\vec{R})\delta(\vec{r}) \notag\\
&+& i W_0(\vec{\sigma}_1+\vec{\sigma_2})[\vec{k}' \times \delta(\vec{r})\vec{k}].
\end{eqnarray}
In the above, $\vec{r}=\vec{r}_1-\vec{r}_2$ is the relative coordinate of the two nucleons, $\vec{R}=(\vec{r}_1+\vec{r}_2)/2$ is their central coordinate with $\rho(\vec{R})$ being the nucleon density there, $\vec{k}=(\nabla_1-\nabla_2)/2i$ is the relative momentum operator and $\vec{k}'$ is its complex conjugate acting on the left, and $P_\sigma=(1+\vec{\sigma}_1 \cdot \vec{\sigma}_2)/2$ is the spin exchange operator, with $\vec{\sigma}_{1(2)}$ being the Pauli matrices acting on nucleon 1(2). While the coefficient of the spin-orbit interaction is fixed at $W_0=133$ MeV fm$^5$, the other nine parameters in the Skyrme interaction $t_0$, $t_1$, $t_2$, $t_3$, $x_0$, $x_1$, $x_2$, $x_3$, and $\alpha$ can be expressed analytically in terms of nine macroscopic quantities, i.e., the saturation density $\rho_0$, the binding energy $E_0$, and the incompressibility $K_0$ of symmetric nuclear matter at $\rho_0$, the isoscalar and isovector nucleon effective mass $m_s^\star$ and $m_v^\star$ in normal nuclear matter, the value $E_{sym}^0$ and the slope parameter $L$ of the symmetry energy at $\rho_0$, and the isoscalar and isovector density gradient coefficient $G_S$ and $G_V$. For more details, we refer the reader to Ref.~\cite{Chen:2010qx}.

As an extension of the above standard SHF EDF, the density-dependent term in the effective interaction [Eq.~(\ref{SHFI})] is replaced by the following term in the KIDS model
\begin{equation}
v^{KIDS}_\rho (\vec{r}_1,\vec{r}_2) = \frac{1}{6}\sum_{i=1}^{3} (t_{3i}+y_{3i}P_\sigma)\rho^{i/3}(\vec{R})\delta(\vec{r}).
\end{equation}
Compared to the standard SHF model, there are three additional coefficients, which allow us to vary three more independent macroscopic quantities, i.e., the skewness EOS parameter $Q_0$ of symmetric nuclear matter, and the curvature parameter $K_{sym}$ as well as the skewness parameter $Q_{sym}$ of the symmetry energy at $\rho_0$. For more details, we refer the reader to Ref.~\cite{PhysRevC.105.044305}.

Based on the effective interaction, the EDF can then be obtained using the Hartree-Fock method, and the single-particle Hamiltonian is obtained using the variational principle, with the Coulomb interaction also explicitly included. Solving the Schr\"odinger equation gives the wave functions of constituent neutrons and protons and thus their density distributions, and the neutron-skin thickness can then be obtained from the difference in the root-mean-square radii of neutrons and protons. For details of this standard procedure, we refer the reader to Ref.~\cite{Vautherin:1971aw}. In the present work, we use Reinhard's code described in Ref.~\cite{Reinhard1991} for the standard SHF model, and a modified one for the KIDS model.

For the RMF model, we take the following Lagrangian form
\begin{equation}\label{lrmf}
\mathcal{L} =  \mathcal{L}_{nm} + \mathcal{L}_\sigma + \mathcal{L}_\omega + \mathcal{L}_\rho + \mathcal{L}_{\omega\rho},
\end{equation}
with
\begin{eqnarray}
\mathcal{L}_{nm} &=& \bar{\psi} (i \gamma^\mu \partial_\mu - m) \psi + g_\sigma \sigma \bar{\psi} \psi - g_\omega \bar{\psi} \gamma^\mu \omega_\mu \psi \notag\\
&-& \frac{g_\rho}{2} \bar{\psi}\gamma^\mu \vec{\rho}_\mu \vec{\tau} \psi, \notag\\
\mathcal{L}_\sigma &=& \frac{1}{2} (\partial^\mu \sigma \partial_\mu \sigma -m_\sigma^2\sigma^2) - \frac{A}{3} \sigma^3 - \frac{B}{4} \sigma^4, \notag\\
\mathcal{L}_\omega &=& -\frac{1}{4} F^{\mu\nu} F_{\mu\nu} + \frac{1}{2} m_\omega^2 \omega_\mu \omega^\mu + \frac{C}{4} (g_\omega^2 \omega_\mu \omega^\mu)^2, \notag\\
\mathcal{L}_\rho &=& -\frac{1}{4} \vec{B}^{\mu\nu} \vec{B}_{\mu\nu} + \frac{1}{2} m_\rho^2 \vec{\rho}_\mu \vec{\rho}^\mu,  \notag\\
\mathcal{L}_{\omega\rho} &=& \frac{1}{2} \alpha_3^\prime g_\omega^2 g_\rho^2 \omega_\mu \omega^\mu \vec{\rho}_\mu \vec{\rho}^\mu. \notag
\end{eqnarray}
In the above, $\mathcal{L}_{nm}$ represents the contribution from the kinetic part of nucleons as well as its coupling to $\sigma$, $\omega$, and $\rho$ mesons, with $\psi$, $\sigma$, $\omega_\mu$, and $\vec{\rho}_\mu$ being the fields of nucleons and corresponding mesons, where $g_\rho$, $g_\omega$, and $g_\rho$ are the corresponding coupling constants, and $\vec{\tau}$ represents the Pauli matrices in isospin space. $\mathcal{L}_\sigma$, $\mathcal{L}_\omega$, and $\mathcal{L}_\rho$ contain free and self-interacting terms of $\sigma$, $\omega$, and $\rho$ mesons, respectively, and $\mathcal{L}_{\omega\rho}$ represents the crossed interaction between $\omega$ and $\rho$ mesons. The antisymmetric field tensors $F_{\mu\nu}$ and $\vec{B}_{\mu\nu}$ are defined as $F_{\mu\nu}=\partial_\nu \omega_\mu - \partial_\mu \omega_\nu$ and $\vec{B}_{\mu\nu}=\partial_\nu \vec{\rho}_\mu - \partial_\mu \vec{\rho}_\nu - g_\rho (\vec{\rho}_\mu \times \vec{\rho}_\nu)$. For a given $C$, the six independent parameters $g_\sigma^2/m_\sigma^2$, $g_\omega^2/m_\omega^2$, $g_\rho^2/m_\rho^2$, $A$, $B$, and $\alpha_3^\prime$ in the RMF model can be expressed inversely in terms of $\rho_0$, $E_0$, $K_0$, $E_{sym}^0$, $L$, and $m_s^\star$ as shown in Ref.~\cite{PhysRevC.90.044305}. The value of $C$ can then be used to vary independently another macroscopic quantity, and we choose it as $Q_0$ as in Ref.~\cite{Zhou:2023hzu}.

Based on the mean-field approximation, the above fields are treated as classical ones. The Euler-Lagrange equations lead to the Dirac equations for nucleons and the Klein-Gordon equations for mesons, and they are solved in a coupled way to get the distributions of various fields in a nucleus, leading to the neutron and proton density distributions as well as the neutron-skin thickness. The calculation is based on the open source code in Ref.~\cite{POSCHL199775}, after the non-linear self-interacting term for $\omega$ meson and the coupling between $\rho$ and $\omega$ mesons are incorporated.

The neutron-star part is calculated in the following way. We assume that the neutron star from the center to the surface contains the liquid core of uniform neutron star matter, the inner crust consisting of nuclear pasta phase, and the outer crust composed of ion lattice and relativistic electron gas. The neutron star matter is formed of neutrons, protons, electrons, and possibly muons, which are in the $\beta$-equilibrium and charge-neutrality condition, and the EOS is obtained from the EDFs of the standard SHF, KIDS, and RMF models described above. The transition density between the liquid core and the inner crust is self-consistently determined from a thermodynamical approach as detailed in Refs.~\cite{Xu:2009vi,PhysRevC.79.035802}. The EOS of the inner crust is parameterized based on an empirical polytropic relation between the pressure and the energy density ~\cite{PhysRevLett.83.3362,LATTIMER2000121,2001ApJ...550..426L}. For the EOS of the outer crust, we take the BPS EOS and the FMT EOS~\cite{1971ApJ...170..299B,Iida_1997}. Here we note that the crust EOS as well as the core-crust transition density may affect the constraints on the EOS from neutron-star observables (see, e.g., Ref.~\cite{Zhou:2023hzu}). With the EOS at all density regions constructed above, the mass-radius relation of neutron stars can be calculated through the Tolman-Oppenheimer-Volkoff equations.

\begin{table}\small
  \caption{Prior ranges of model parameters in the standard SHF, KIDS, and RMF models for the Bayesian analysis in the present study.}
    \begin{tabular}{|c|c|c|c|c|}
  \hline
    & SHF & KIDS & RMF \\
   \hline
    $K_0$ (MeV) & 220 $-$ 260  & 220 $-$ 260 & 220 $-$ 260  \\
    $Q_0$ (MeV) & - & -800 $-$ 400 & -800 $-$ 400 \\
    $E_{sym}^0$ (MeV) & 28.5 $-$ 34.9  & 28.5 $-$ 34.9  & 28.5 $-$ 34.9 \\
    $L$ (MeV) & 30 $-$ 90 & 30 $-$ 90 & 30 $-$ 90\\
    $K_{sym}$ (MeV) & - & -400 $-$ 100 & -  \\
    $Q_{sym}$ (MeV) & - & -200 $-$ 800 & -  \\
    $m_s^\star/m$ & 0.5 $-$ 0.9 & 0.5 $-$ 0.9 & 0.5 $-$ 0.9 \\
    $m_v^\star/m$ & 0.5 $-$ 0.9 & 0.5 $-$ 0.9 & - \\
   \hline
    \end{tabular}
  \label{T1}
\end{table}

The Bayesian analysis extracts posterior probability distribution functions (PDFs) of model parameters $M(p_1,p_2,p_3,...)$ by comparing results $D(d_1,d_2,d_3,...)$ from theoretical calculations to the experimental data. The numbers of model parameters, which are set as independent macroscopic quantities as described above, are different in the standard SHF, KIDS, and RMF models. In order to carry out a fair comparison for the three models, we have fixed $\rho_0=0.16$ fm$^{-3}$ and $E_0=-16$ MeV for all models, and $G_S=132$ MeV fm$^5$ and $G_V=5$ MeV fm$^5$ for the standard SHF and KIDS models, according to the empirical values of model parameters in Ref.~\cite{Chen:2010qx}.  We choose to vary $p_1=K_0$ uniformly within $220-260$ MeV from studies on isoscalar giant monopole resonances~\cite{PhysRevLett.109.092501,PhysRevC.97.025805,Shlomo2006,Colo:2013yta,Garg:2018uam}, and $p_2=E_{sym}^0$ and $p_3=L$ uniformly within $28.5-34.9$ MeV and $30-90$ MeV, respectively, according to Refs.~\cite{LI2013276,RevModPhys.89.015007}. Higher-order EOS parameters $p_4=K_{sym}$, $p_5=Q_0$, and $p_6=Q_{sym}$, if they can be changed as independent model parameters, are varied uniformly within their prior ranges obtained based on analyses of terrestrial nuclear experiments and EDFs~\cite{Tews_2017,Zhang:2017ncy}. We also vary the non-relativistic isoscalar and isovector p-masses, i.e., $p_7=m_s^\star$ and $p_8=m_v^\star/m$, for the standard SHF and KIDS models, and the isoscalar Dirac effective mass $p_7=m_s^\star/m$ for the RMF model. The prior ranges of model parameters are listed in Table \ref{T1} for different models, and in the Bayesian analysis a random walk is performed in such parameter space. As shown in Ref.~\cite{Zhou:2023hzu}, the real parameter space is smaller for the RMF model when we try to study properties of neutron stars, since there could be no solutions for the field equations at high densities. In the study of finite nuclei, however, we have a larger parameter space for the RMF model compared to that in Ref.~\cite{Zhou:2023hzu}.

For neutron skins, how well the results $d_{i}^{th}$ obtained from the theoretical model with model parameters $p_{i}$ reproduce the experimental data $d_{i}^{exp}$ is described by the likelihood function
\begin{eqnarray}
&&P_{\Delta r_{np}} = \Pi_{i=1,2} \Bigg \{ \frac{1}{2\pi \sigma_i} \exp\left[-\frac{(d^{th}_i-d^{exp}_i)^2}{2\sigma_i^2}\right] \notag\\
&\times& \Theta\left(0.03-\left|\frac{E_i^{th}-E_i^{exp}}{E_i^{exp}}\right|\right)\Theta\left(0.03-\left|\frac{r_i^{th}-r_i^{exp}}{r_i^{exp}}\right|\right)\Bigg\}. \label{llh_drnp}
\end{eqnarray}
Here $i=1$ and 2 represent data of $^{208}$Pb and $^{48}$Ca, respectively. We choose the neutron-skin thickness data $\Delta r_{np} = 0.283\pm0.071$ fm for $^{208}$Pb from PREX~\cite{PREX:2021umo} and $\Delta r_{np}= 0.121\pm0.035$ fm for $^{48}$Ca from CREX~\cite{CREX:2022kgg}, so $d^{exp}_{1,2}$ and $\sigma_{1,2}$ in Eq.~(\ref{llh_drnp}) are chosen to be the mean values and $1\sigma$ errors, respectively. While each theoretical model can reproduce the experimental data of the binding energy $E_i^{exp}$ and the charge radius $r_i^{exp}$ taken from Refs.~\cite{Audi:2003zz,Angeli:2004kvy} at a higher accuracy, here we allow a rather extensive error $3\%$ so that the extracted constraints on the EOS are from neutron-skin data rather than from $E_i^{exp}$ and $r_i^{exp}$.

\begin{figure}[!h]
\includegraphics[width=1.0\linewidth]{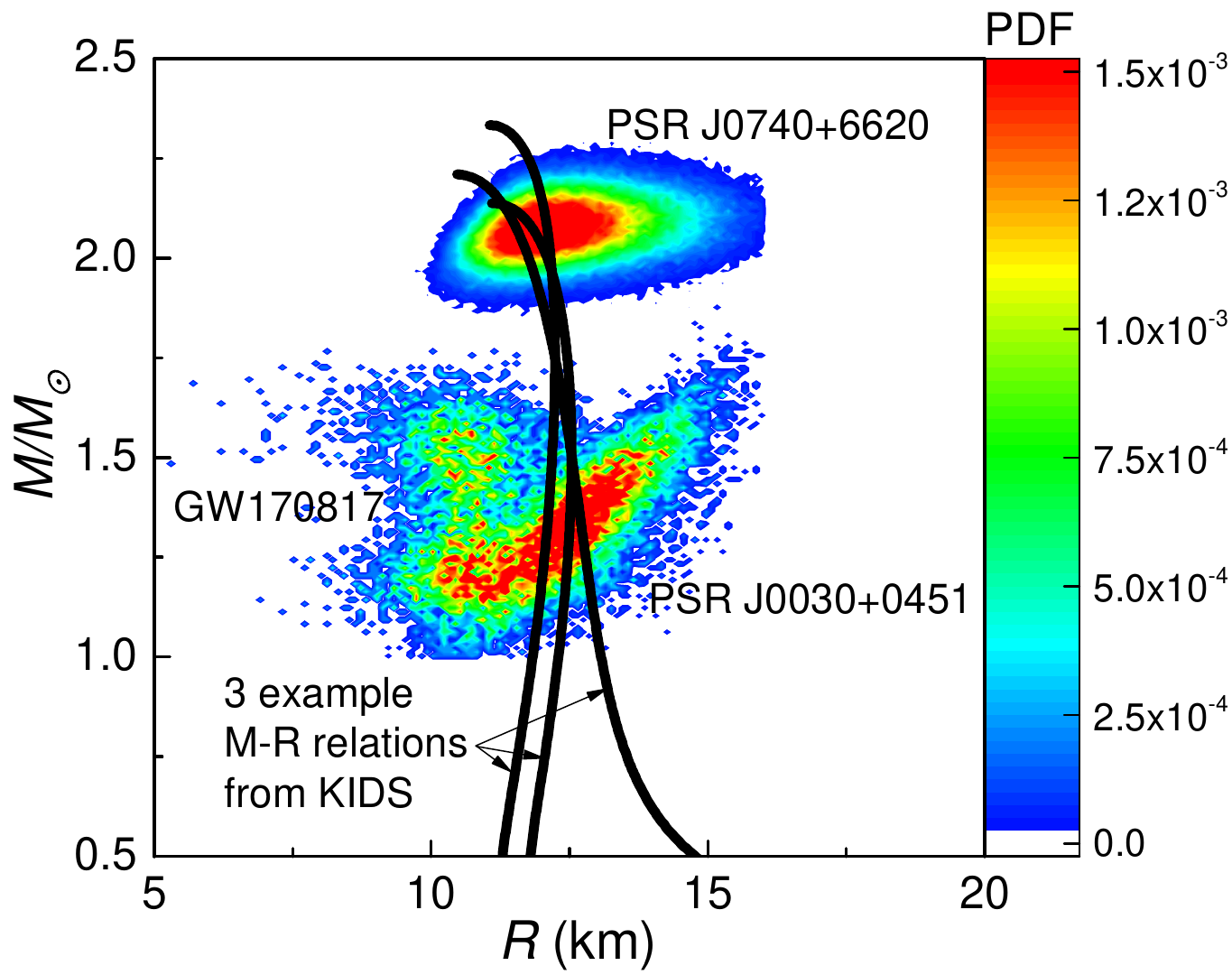}
\caption{\label{MR} Two-dimensional PDFs from the sampling data of the GW170817 event as well as the PSR J0030+0451 and the PSR J0740+6620 in the mass-radius ($M-R$) plane. Three representative highly-favored $M-R$ relations from the KIDS model, which pass through the most probable regions of the PDFs, are plotted for illustration.}
\end{figure}

From the GW170817 event measured by the LIGO Scientific Collaboration and Virgo Collaboration, the radii of the binary stars were measured to be $R_1=10.8^{+2.0}_{-1.7}$ km and $R_2=10.7^{+2.1}_{-1.5}$ km from the EOS-insensitive-relation analysis, with the masses $M_1$ within $[1.36, 1.62]M_{\odot}$ and $M_2$ within $[1.15, 1.36]M_{\odot}$, where $M_{\odot}$ is the solar mass, and a tidal deformability $\Lambda_{1.4}=190^{+390}_{-120}$ for canonical neutron stars is also inferred from the GW170817 data at the $90\%$ confidence level~\cite{LIGOScientific:2018cki}. More recently, the mass and the radius of the pulsar PSR J0030+0451 were obtained to be $1.34^{+0.15}_{-0.16}M_{\odot}$ and $12.71^{+1.14}_{-1.19}$ km~\cite{Riley:2019yda}, respectively, and those of the pulsar PSR J0740+6620 were constrained to be $2.072^{+0.067}_{-0.066}M_{\odot}$ and $12.39^{+1.30}_{-0.98}$ km~\cite{Riley_2021}, respectively, at the $68\%$ confidence level by NICER. In the present study, we construct the two-dimensional PDFs in the $M-R$ plane from the sampling data of the GW170817 event as well as the PSR J0030+0451 and the PSR J0740+6620, as those in Fig. 3 of Ref.~\cite{LIGOScientific:2018cki}, Fig. 20 of Ref.~\cite{Riley:2019yda}, and Fig. 7 of Ref.~\cite{Riley_2021}, respectively. The resulting PDFs for the three events are shown in Fig.~\ref{MR}, where three representative highly-favored $M-R$ relations from the KIDS model, which pass through the most probable regions of the PDFs, are plotted for illustration. To be quantitative, we have mapped the two-dimensional PDFs $f_n(M,R)$ in the $M-R$ plane to $N_M\times N_R$ lattices, with $n=1$, 2, and 3 representing the three astrophysical events, i.e., GW170817, PSR J0030+0451, and PSR J0740+6620, and $f_n(M,R)$ is normalized for each $n$. The likelihood function describing how well the resulting $M-R$ curve $M(R)$ reproduces the data is calculated by summing the values of $f_n(M,R)$ in the lattices along the trajectory of $M(R)$ and multiplying those for the three events, i.e.,
\begin{eqnarray}
&&P_{MR} = \Pi^3_{n=1} \left[\sum_{j \in \tilde{M}(R)} f_n(M_j,R_j) \right]. \label{llh_MR}
\end{eqnarray}
We use $N_M=200$ and $N_R=150$ in the present analysis, and their values may affect the resolution but may not affect the final results by much. In the summation of Eq.~(\ref{llh_MR}), we have also subtracted parts of the $M(R)$ curve which represent unstable neutron stars or contain neutron-star matter that violates the causality condition, and the modified trajectory is expressed as $\tilde{M}(R)$.

The total likelihood function is $P=P_{\Delta r_{np}} \times P_{MR}$. According to the Bayes' theorem, the posterior PDF is the product of the likelihood function and the prior PDF with normalization. In the real calculation, the resulting posterior PDFs of EOS parameters from neutron-skin data or neutron-star data alone can be taken as the prior PDFs used for the second-round calculation, to achieve the final posterior PDFs from both neutron-skin and neutron-star data. For the algorithm of the Bayesian analysis, a Markov-Chain Monte Carlo approach using the Metropolis-Hastings algorithm is employed to reach an equilibrium distribution, with the relaxation process subtracted in the final analysis.

\section{Results and discussions}
\label{sec:results}

\begin{figure}[!h]
\includegraphics[width=1.0\linewidth]{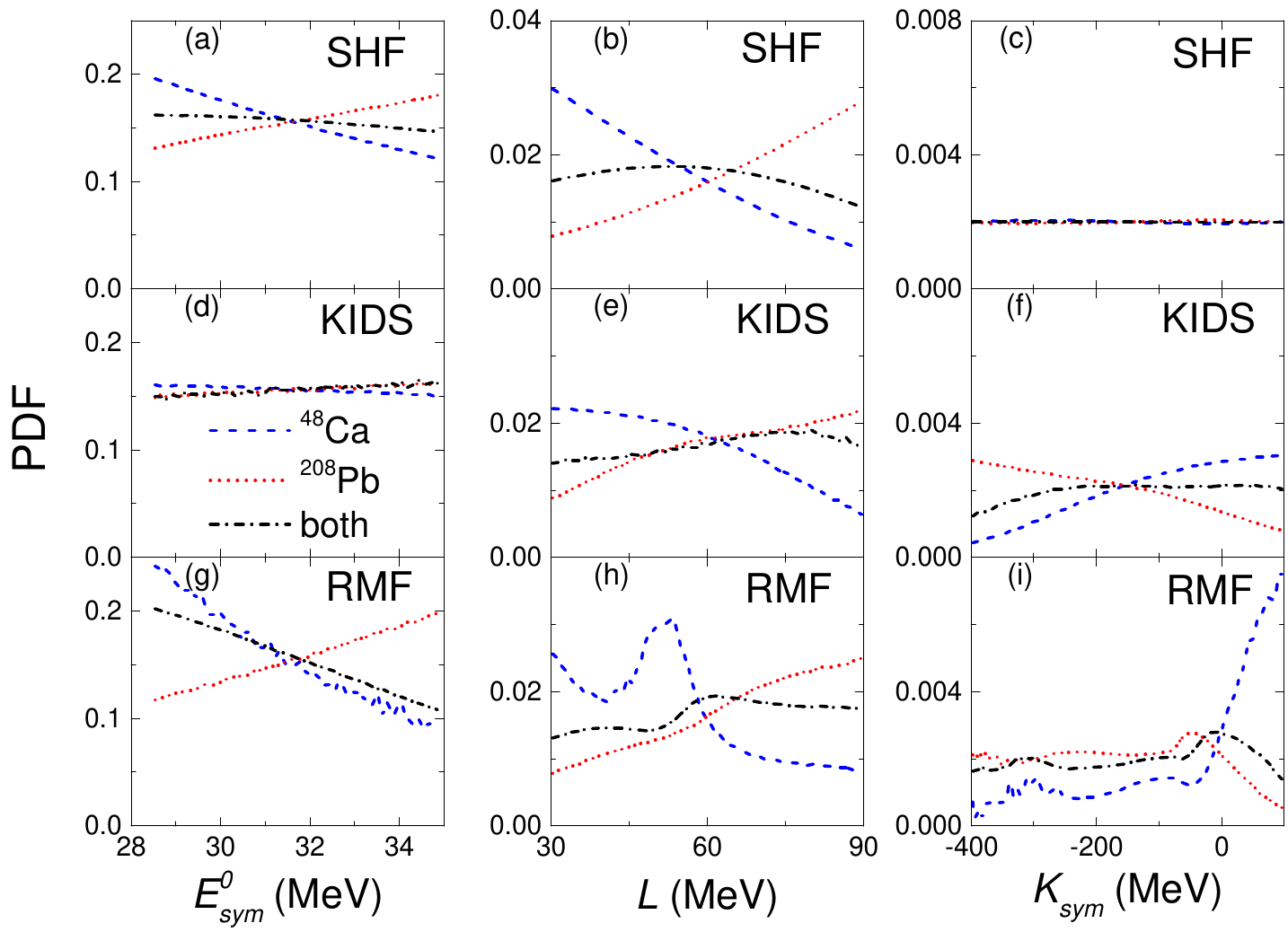}
\caption{\label{fig1} Posterior PDFs of $E_{sym}^0$ (left), $L$ (middle), and $K_{sym}$ (right) from the neutron-skin thickness data of $^{48}$Ca, $^{208}$Pb, and both, based on the standard SHF [(a)-(c)], KIDS [(d)-(f)], and RMF [(g)-(i)] models. }
\end{figure}

We start by comparing the posterior PDFs of different symmetry energy parameters from the constraints of the neutron-skin thickness $\Delta r_{np}$ data for $^{48}$Ca and $^{208}$Pb based on different effective nuclear interactions in Fig.~\ref{fig1}. While $E_{sym}^0$ and $L$ are anti-correlated from the constraint of $\Delta r_{np}$~\cite{Xu:2020fdc} based on the standard SHF model, the detailed behaviors of their posterior PDFs depend on their prior ranges (see, e.g., Fig.~4 in Ref.~\cite{PhysRevC.105.044305}), and a small (large) experimental value of $\Delta r_{np}$ for $^{48}$Ca ($^{208}$Pb) favors both small (large) $E_{sym}^0$ and $L$. In the standard SHF model where $K_{sym}$ can't be varied independently, the $\Delta r_{np}$ can't constrain $K_{sym}$. In the KIDS model where $K_{sym}$ can be varied as an independent model parameter, the constraint on $E_{sym}^0$ becomes weaker while an opposite constraint on $K_{sym}$ compared to that on $L$ is observed. The latter is understandable since parameters of the symmetry energy at different orders compensate for each other. In the RMF model, some abnormal behaviors are observed, especially for the posterior PDFs of $L$ and $K_{sym}$, and the explanations can be found in Appendix \ref{appendix}. If we adopt both constraints of $\Delta r_{np}$ for $^{48}$Ca and $^{208}$Pb, the resulting posterior PDFs are roughly the average of the PDFs from only $^{48}$Ca or $^{208}$Pb, which actually favor opposite trends of $E_{sym}(\rho)$ and corresponding parameters.

\begin{figure}[!h]
\includegraphics[width=1.0\linewidth]{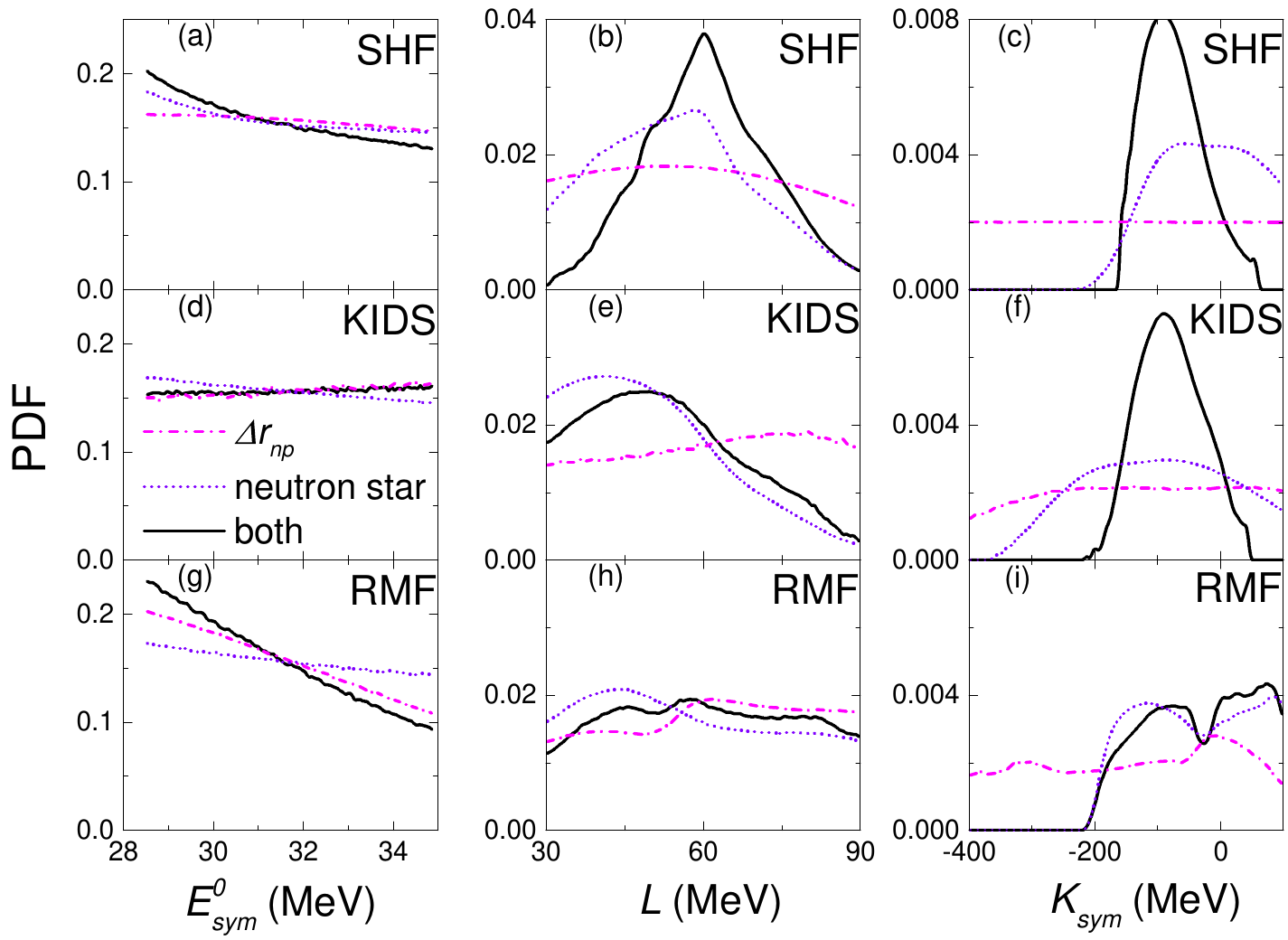}
\caption{\label{fig2} Posterior PDFs of $E_{sym}^0$ (left), $L$ (middle), and $K_{sym}$ (right) from the neutron-skin thickness data, the neutron-star observables, and both data sets, based on the standard SHF [(a)-(c)], KIDS [(d)-(f)], and RMF [(g)-(i)] models. }
\end{figure}

The posterior PDFs of symmetry energy parameters from both constraints of $\Delta r_{np}$ for $^{48}$Ca and $^{208}$Pb are compared with those from neutron-star observables~\cite{Zhou:2023hzu} in Fig.~\ref{fig2}. It is seen that neutron-star observables generally favor a relatively smaller $L$ but a larger $K_{sym}$, from respectively the constraints of the radii for intermediate- and heavy-mass neutron stars from the three astrophysical events, compared to the constraints from the $\Delta r_{np}$ for both $^{48}$Ca and $^{208}$Pb. It is noteworthy that the constraints on $K_{sym}$ mostly come from neutron-star obserables rather than from $\Delta r_{np}$. The constraining powers on the $E_{sym}$ parameters generally become enhanced after the data of neutron-star observables are taken into account, especially for the standard SHF and KIDS models. The dip region of the $K_{sym}$ PDF in the RMF model corresponds to the parameter space with smaller $m_s^\star$ and larger $L$, which can not access a two-solar-mass neutron star, while other combinations of $m_s^\star$ and $L$ which lead to the rest regions of $K_{sym}$ can explain better the astrophysical data. With the posterior PDFs shown in Figs.~\ref{fig1} and \ref{fig2}, we do see model dependence on the constraints of the $E_{sym}$ parameters based on the same data set, depending on the EDF form and the number of independent model parameters.

\begin{figure*}[!h]
\includegraphics[width=0.22\linewidth]{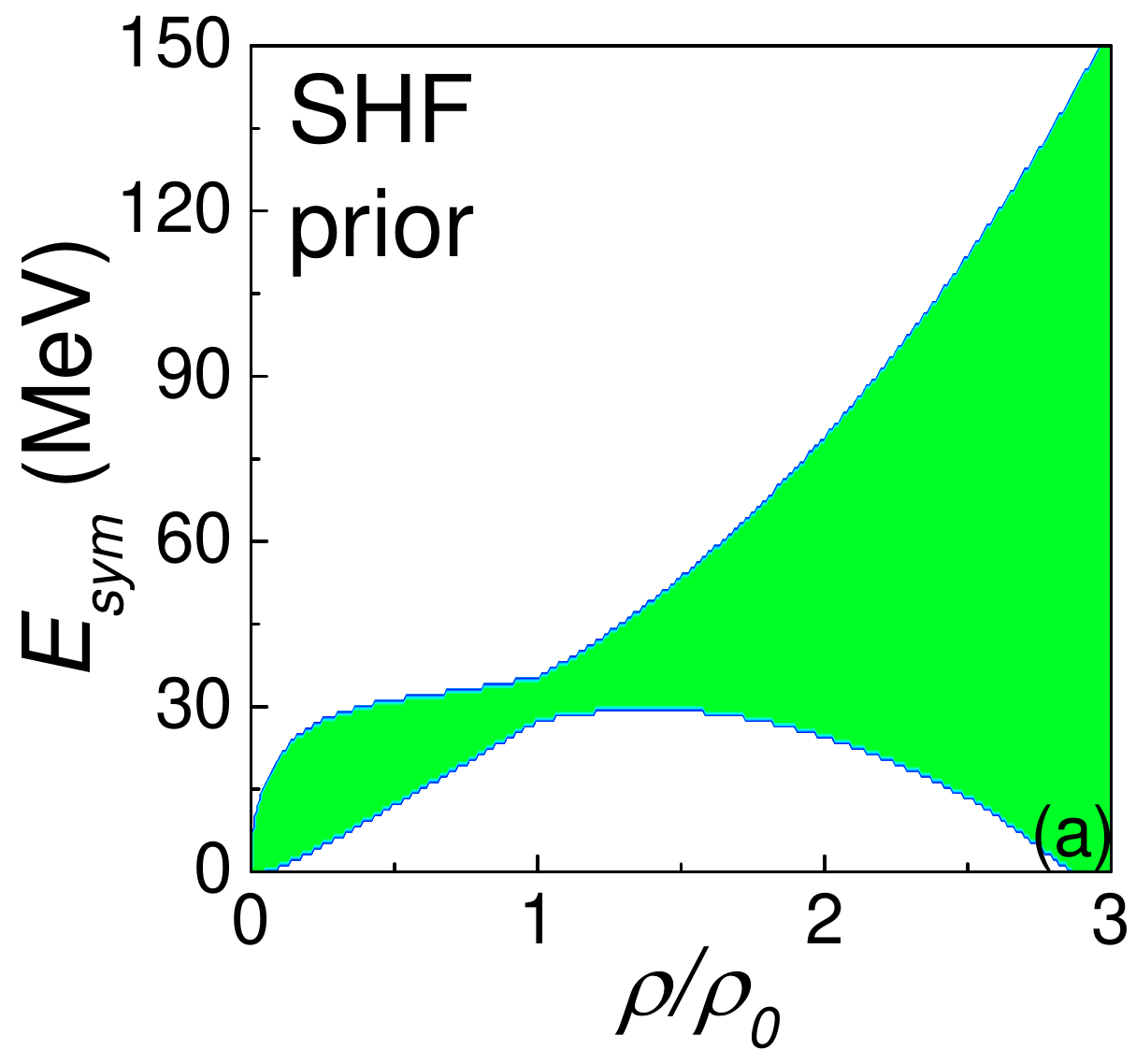}
\includegraphics[width=0.22\linewidth]{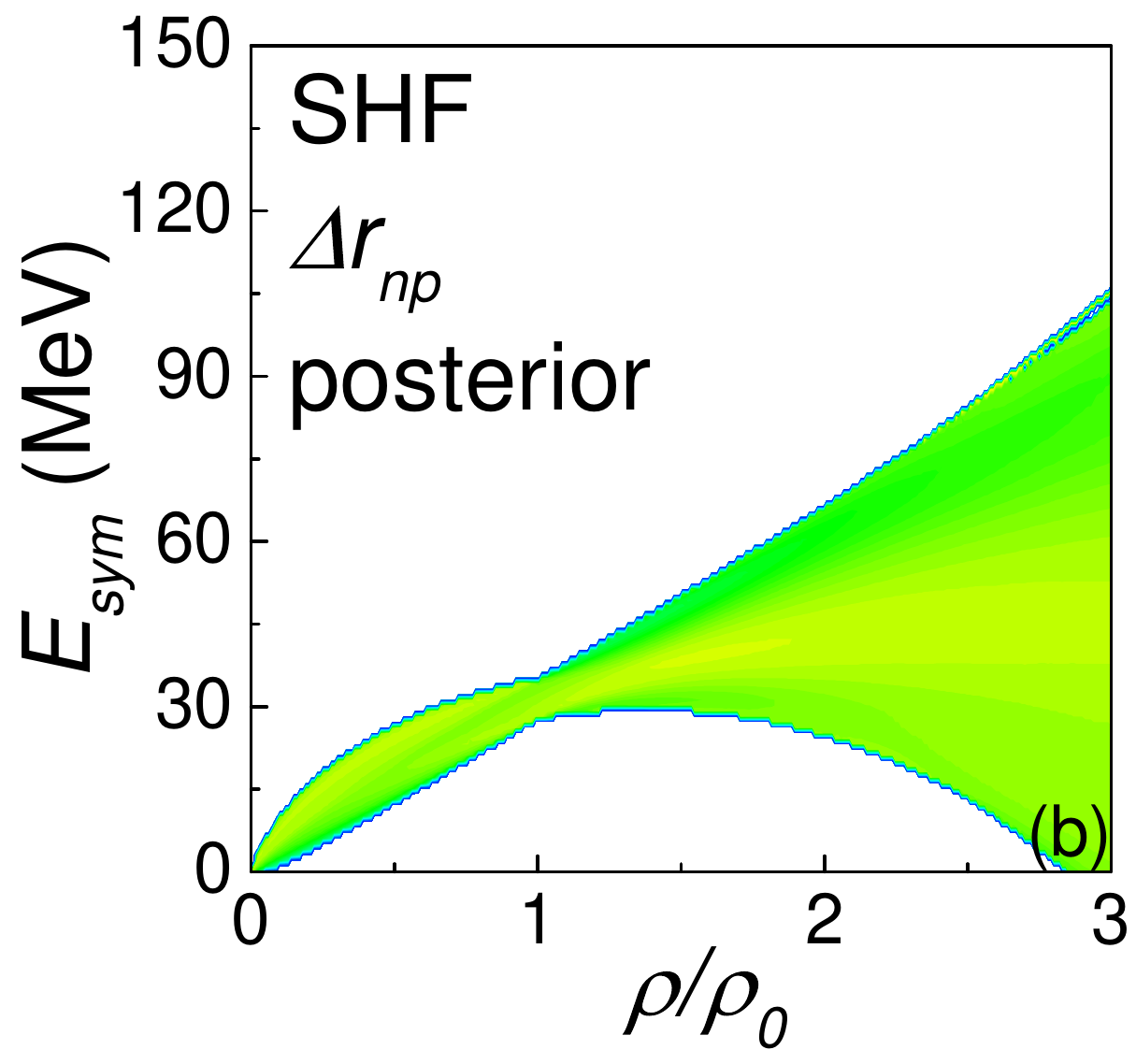}
\includegraphics[width=0.22\linewidth]{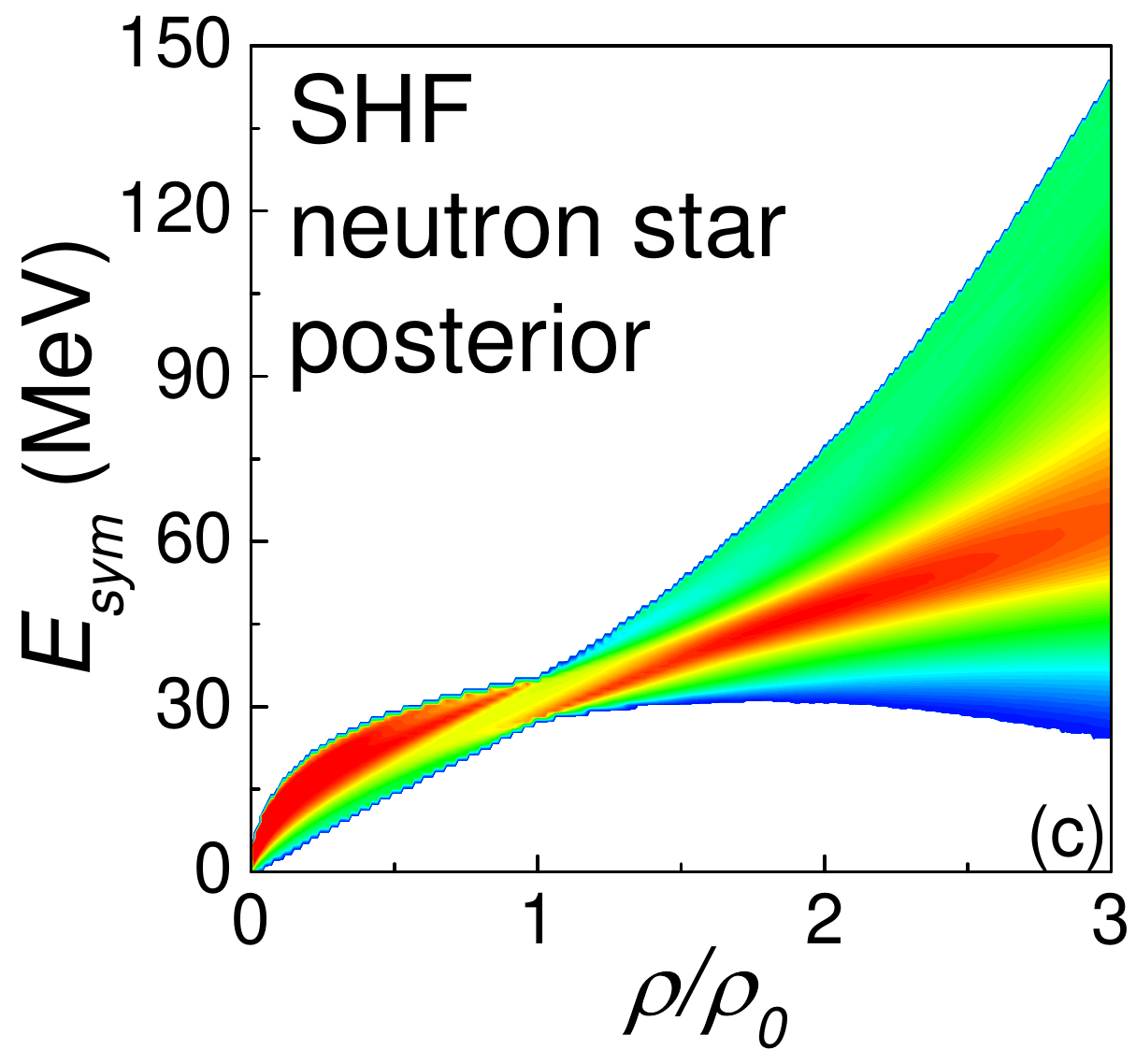}
\includegraphics[width=0.27\linewidth]{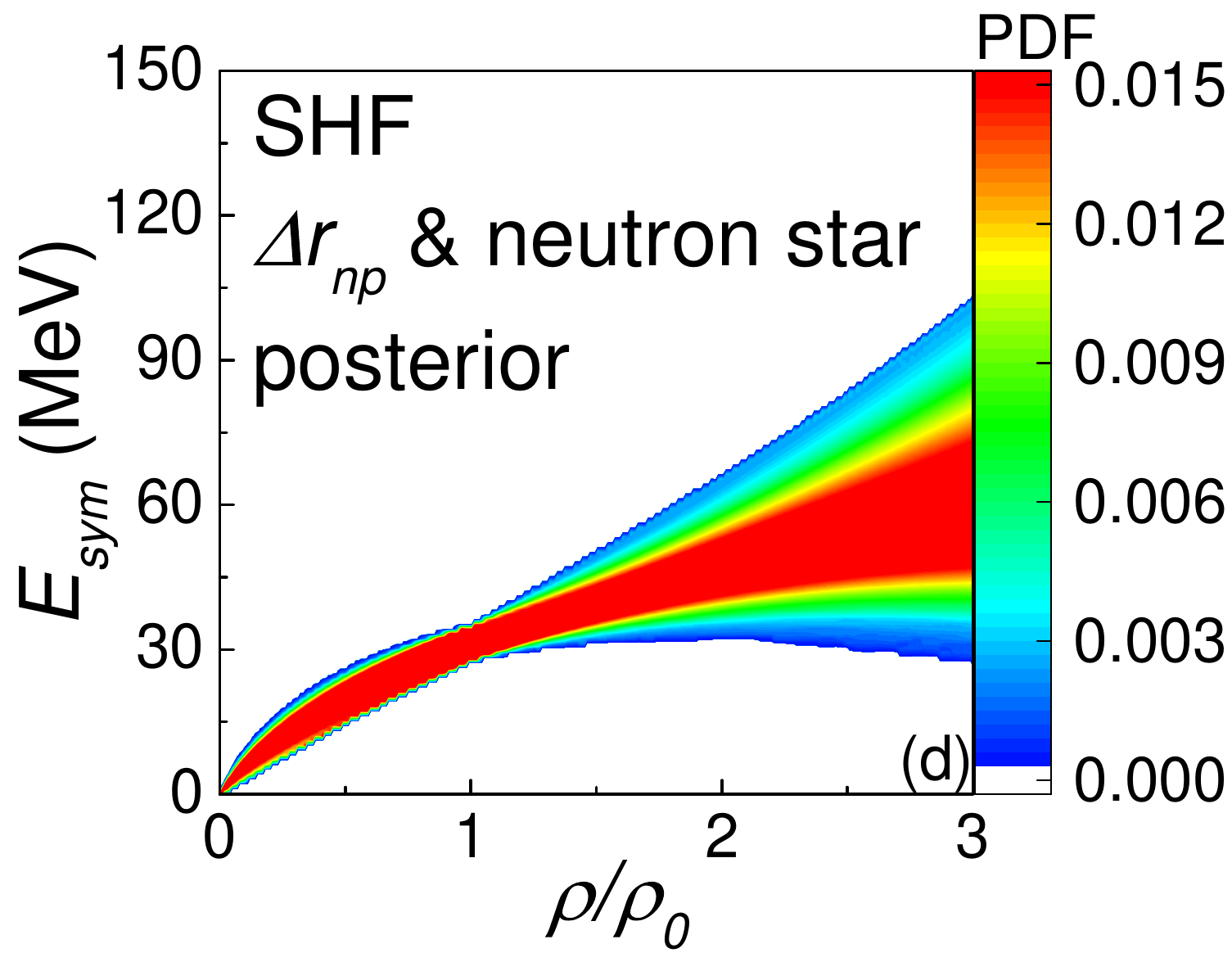}\\
\includegraphics[width=0.22\linewidth]{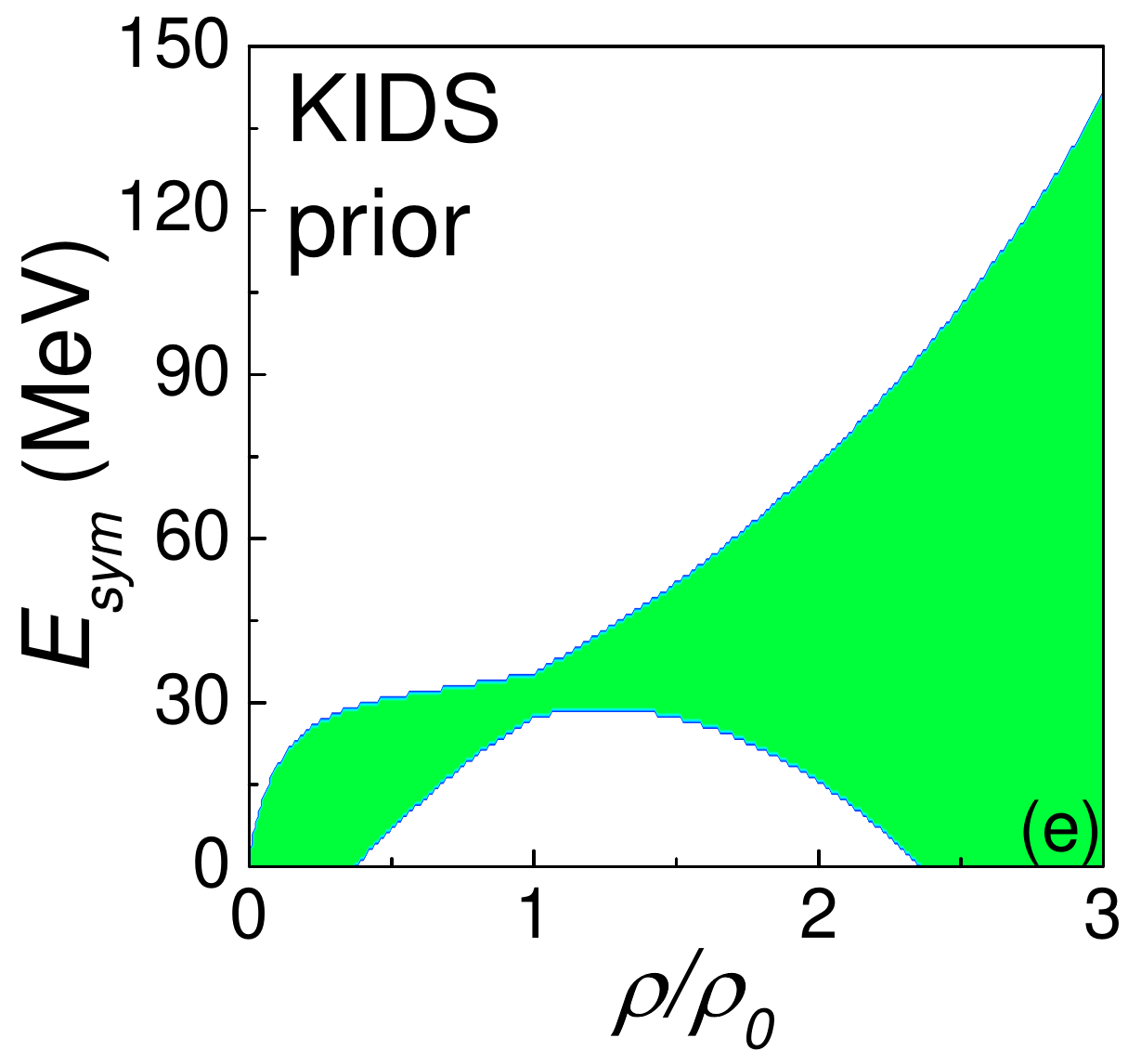}
\includegraphics[width=0.22\linewidth]{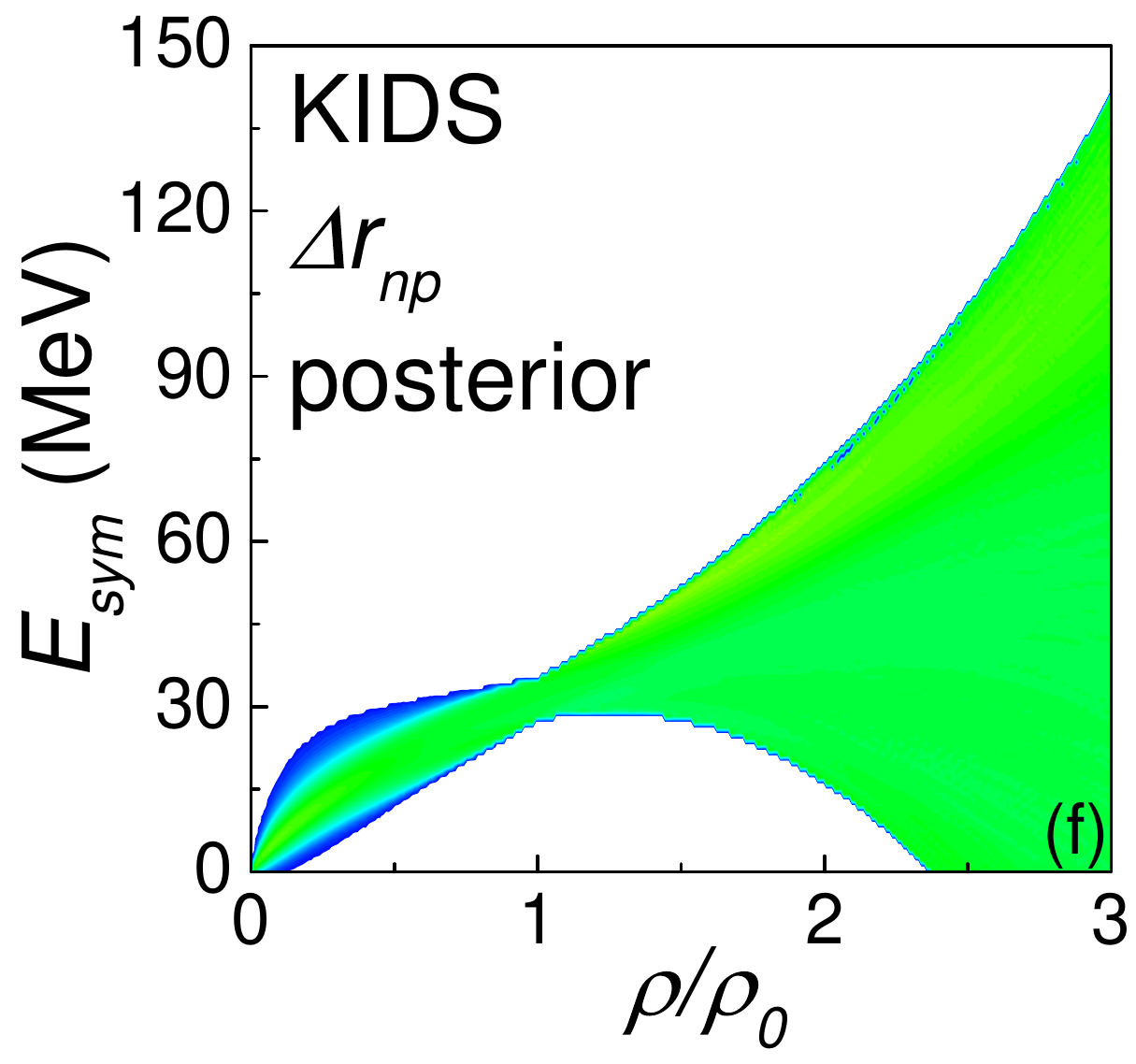}
\includegraphics[width=0.22\linewidth]{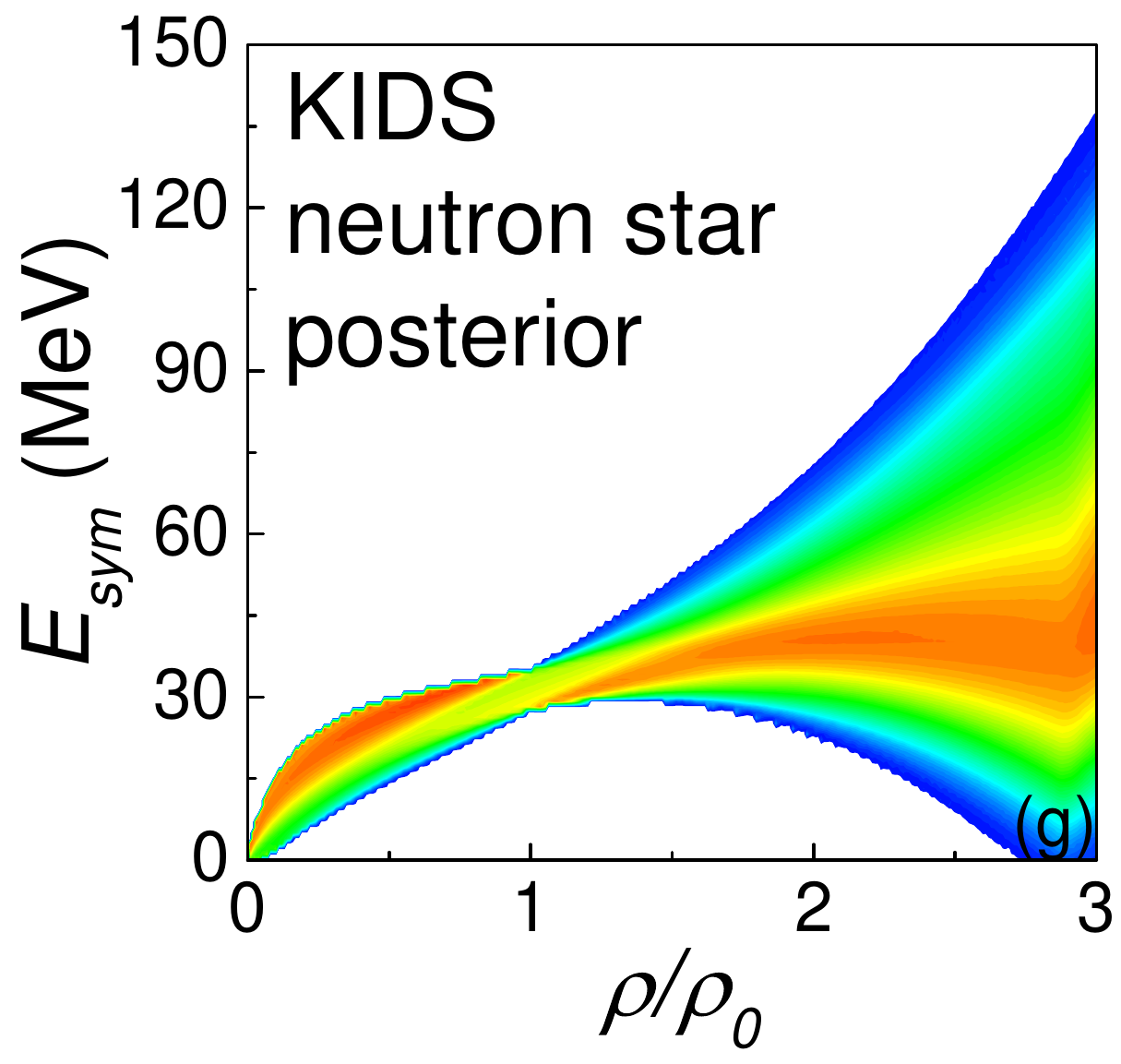}
\includegraphics[width=0.27\linewidth]{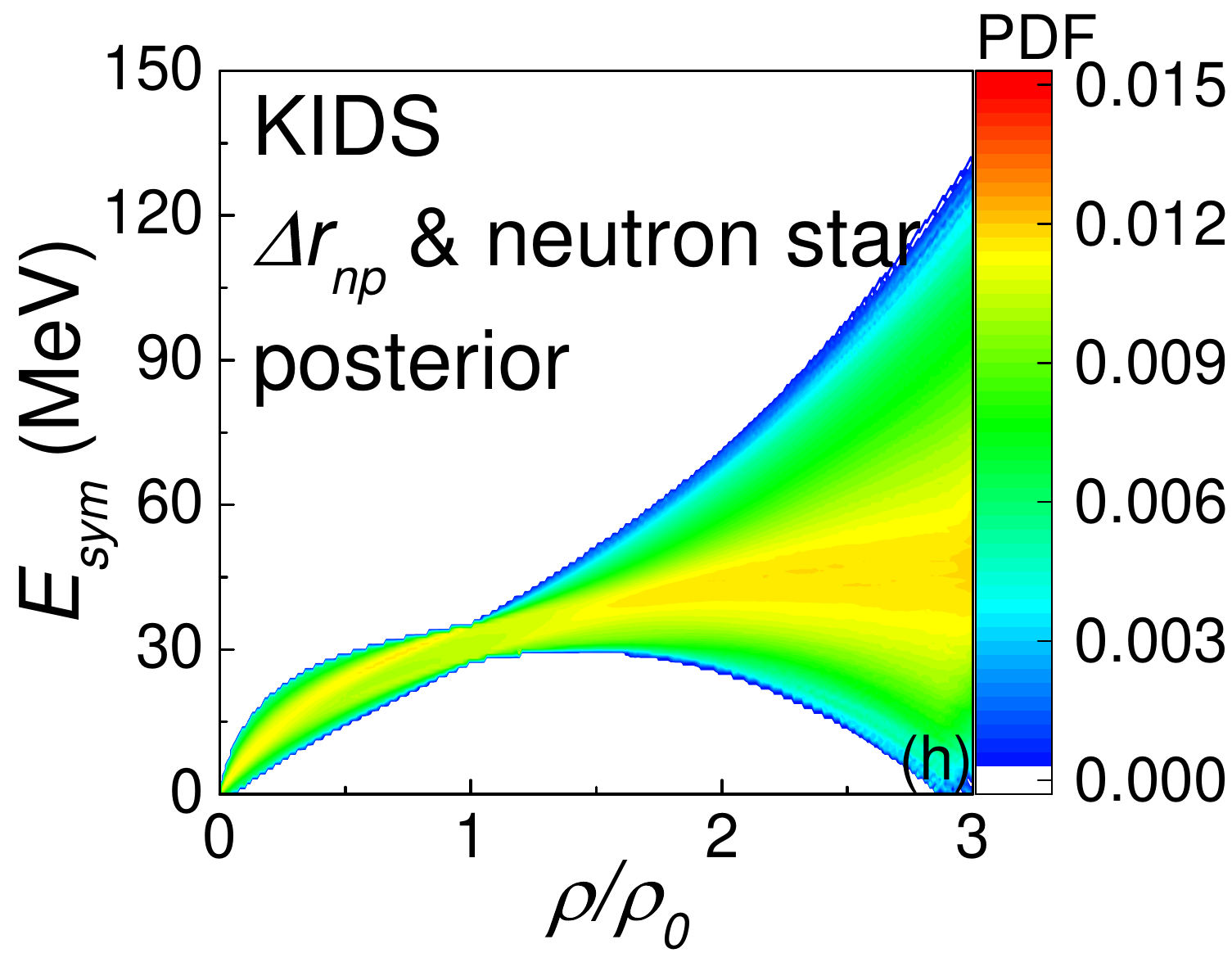}\\
\includegraphics[width=0.23\linewidth]{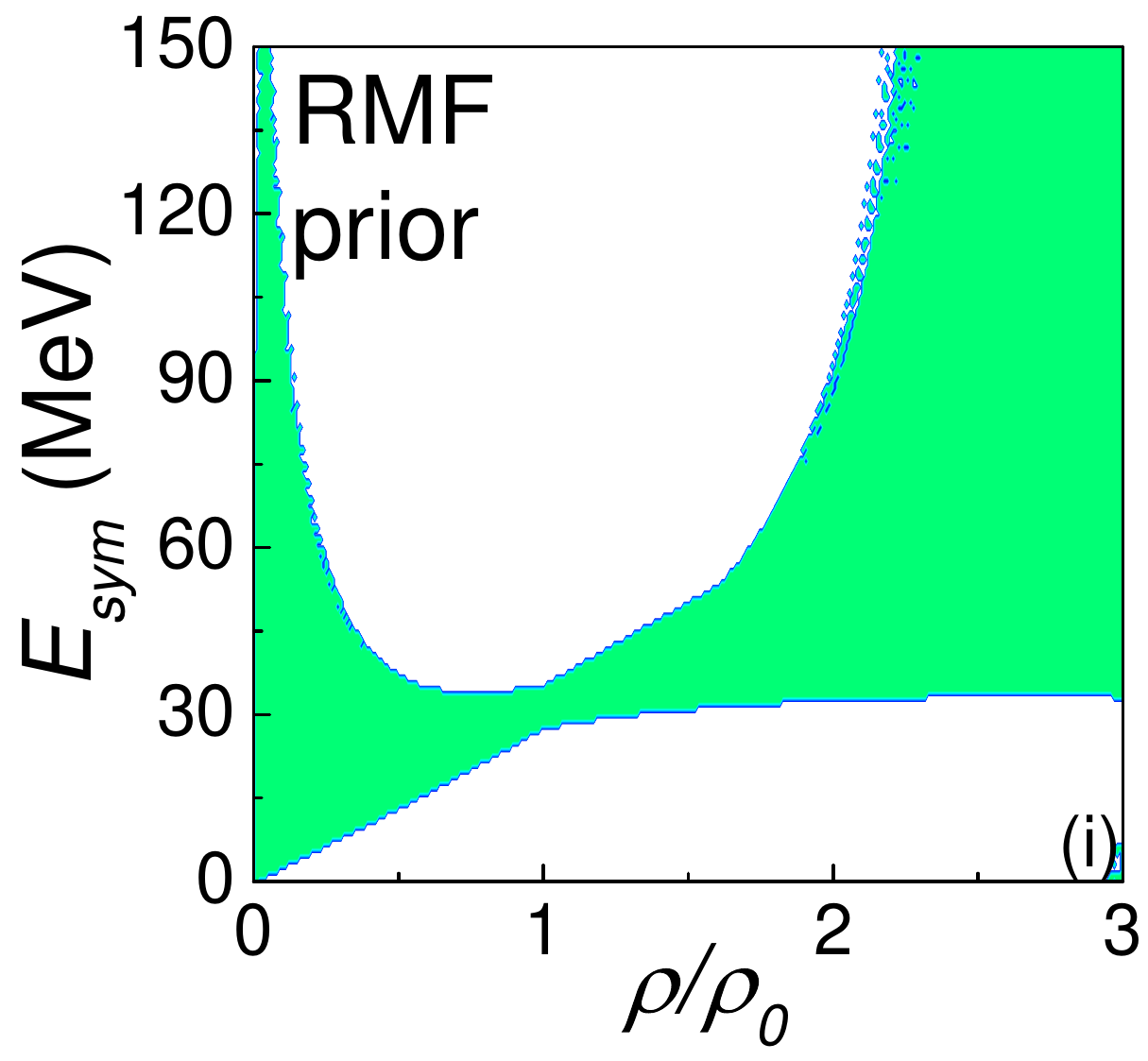}
\includegraphics[width=0.22\linewidth]{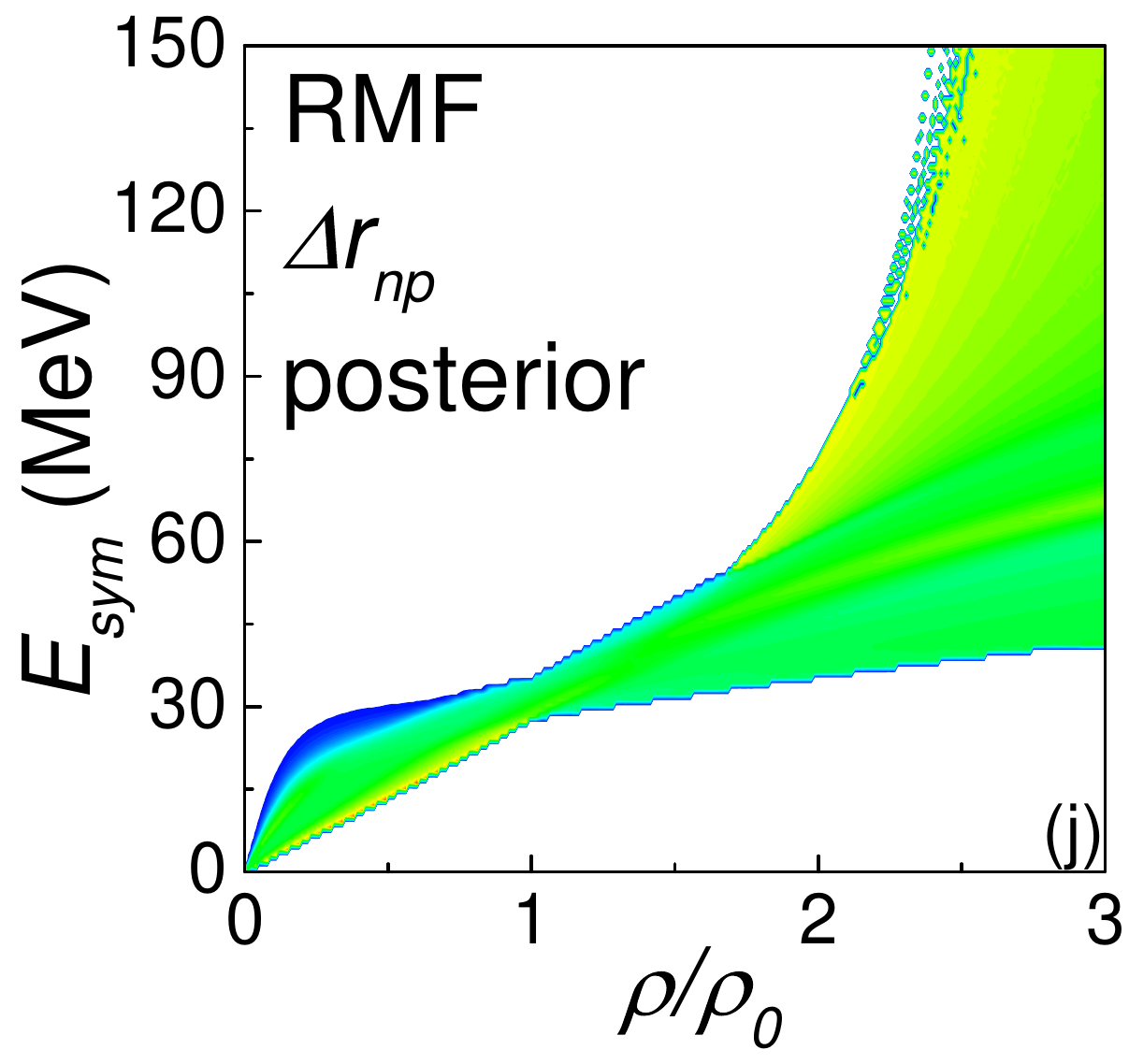}
\includegraphics[width=0.22\linewidth]{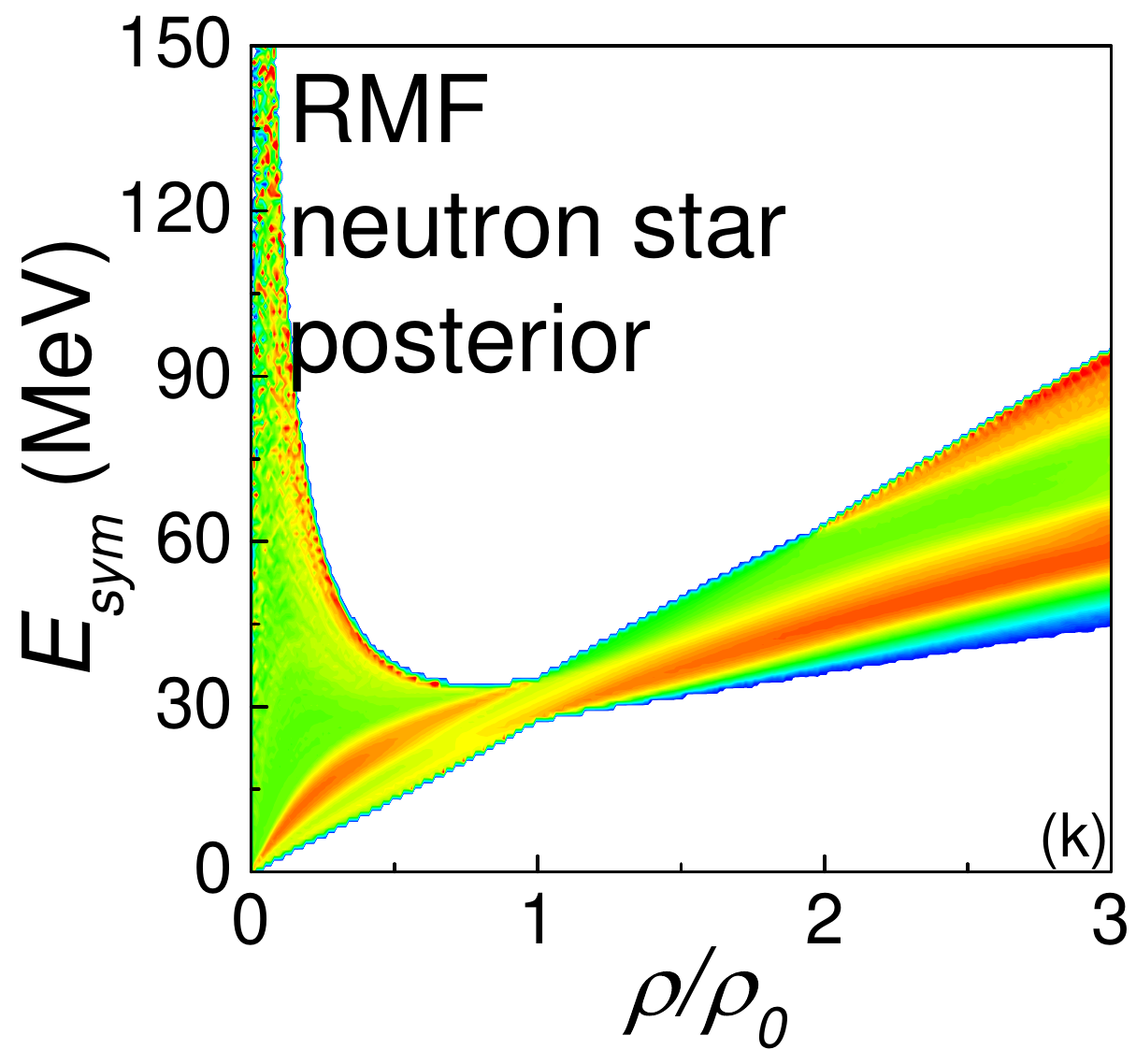}
\includegraphics[width=0.27\linewidth]{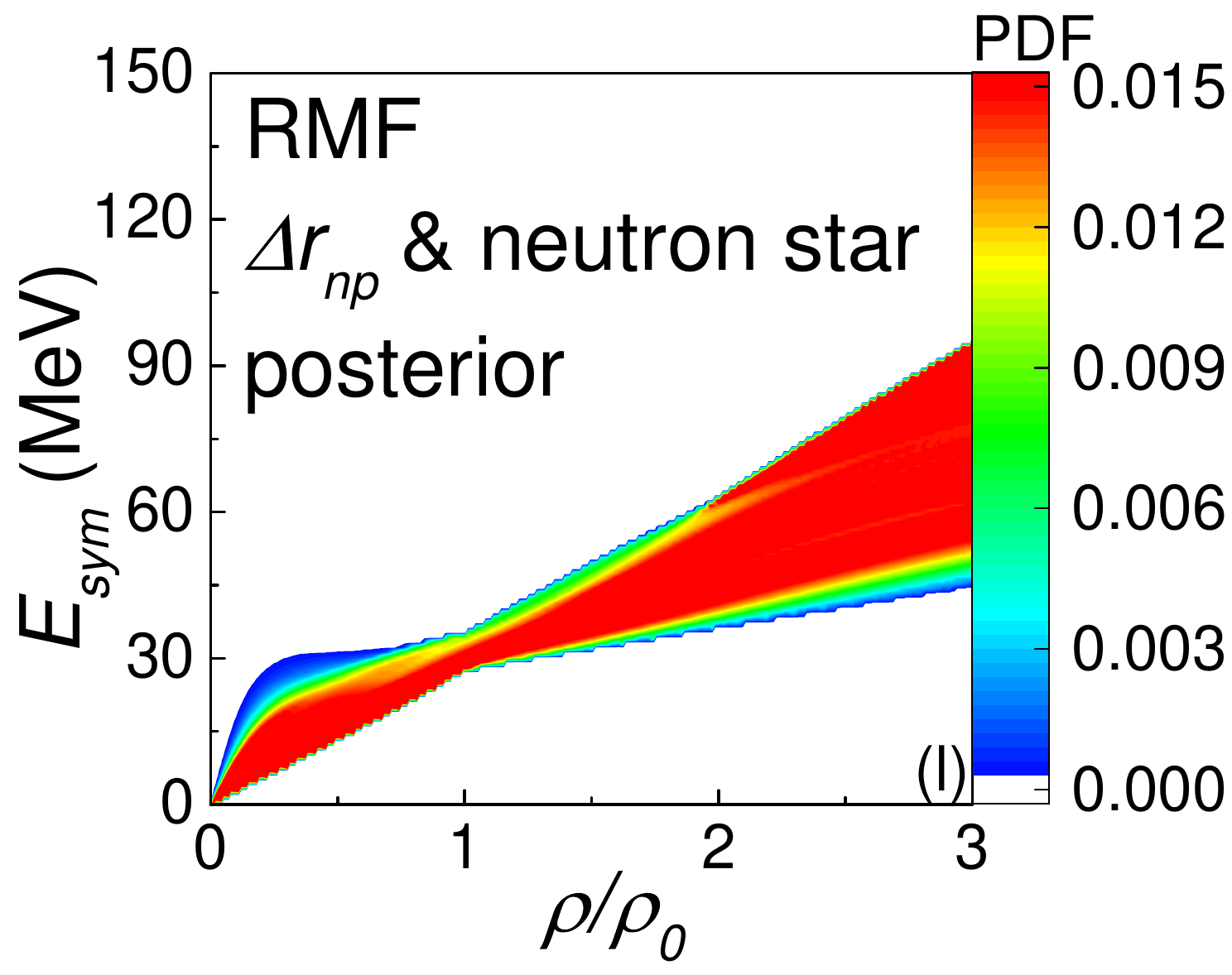}
\caption{\label{fig3} Prior PDFs (first column) and posterior PDFs of the nuclear symmetry energy from the neutron-skin thickness data (second column), neutron-star observables (third column), and both data sets (fourth column), based on the standard SHF [(a)-(d)], KIDS [(e)-(h)], and RMF [(i)-(l)] models. }
\end{figure*}

We compare the density dependence of the symmetry energy with prior distributions of model parameters and those from the constraints of $\Delta r_{np}$ data, neutron-star observables, and both data sets in the density range from $0$ to $3\rho_0$ in Fig.~\ref{fig3}. The prior ranges of the symmetry energy are similar for the standard SHF and KIDS model, by adopting the prior ranges of independent model parameters as listed in Table ~\ref{T1}, except that KIDS allows an even softer symmetry energy. For the RMF model, as mentioned above, the larger parameter space for the study of finite nuclei leads to a larger prior range of $E_{sym}(\rho)$ compared to that in the neutron-star study~\cite{Zhou:2023hzu}. The $\Delta r_{np}$ data put some constraints on the symmetry energy at subsaturation densities for all three models, especially for the RMF model by ruling out too large and unphysical $E_{sym}(\rho)$ at low densities, while the constraining power at high densities is weak. The neutron-star observables constrain appreciably the symmetry energy at suprasaturation densities, while the constraining power at subsaturation densities depends on the corresponding EDF form. The favored stiff $E_{sym}$ in the RMF model from the constraint of neutron-star observables is consistent with the behavior of the $K_{sym}$ PDF shown in Fig.~\ref{fig2}(i). Combining both constraints from $\Delta r_{np}$ and neutron-star observables, the symmetry energies from low to high densities are nicely constrained. The constraining power is stronger for the standard SHF model and the RMF model, and weaker for the KIDS model, due to a larger number of independent model parameters in the KIDS model.

\begin{figure}[!h]
\includegraphics[width=0.8\linewidth]{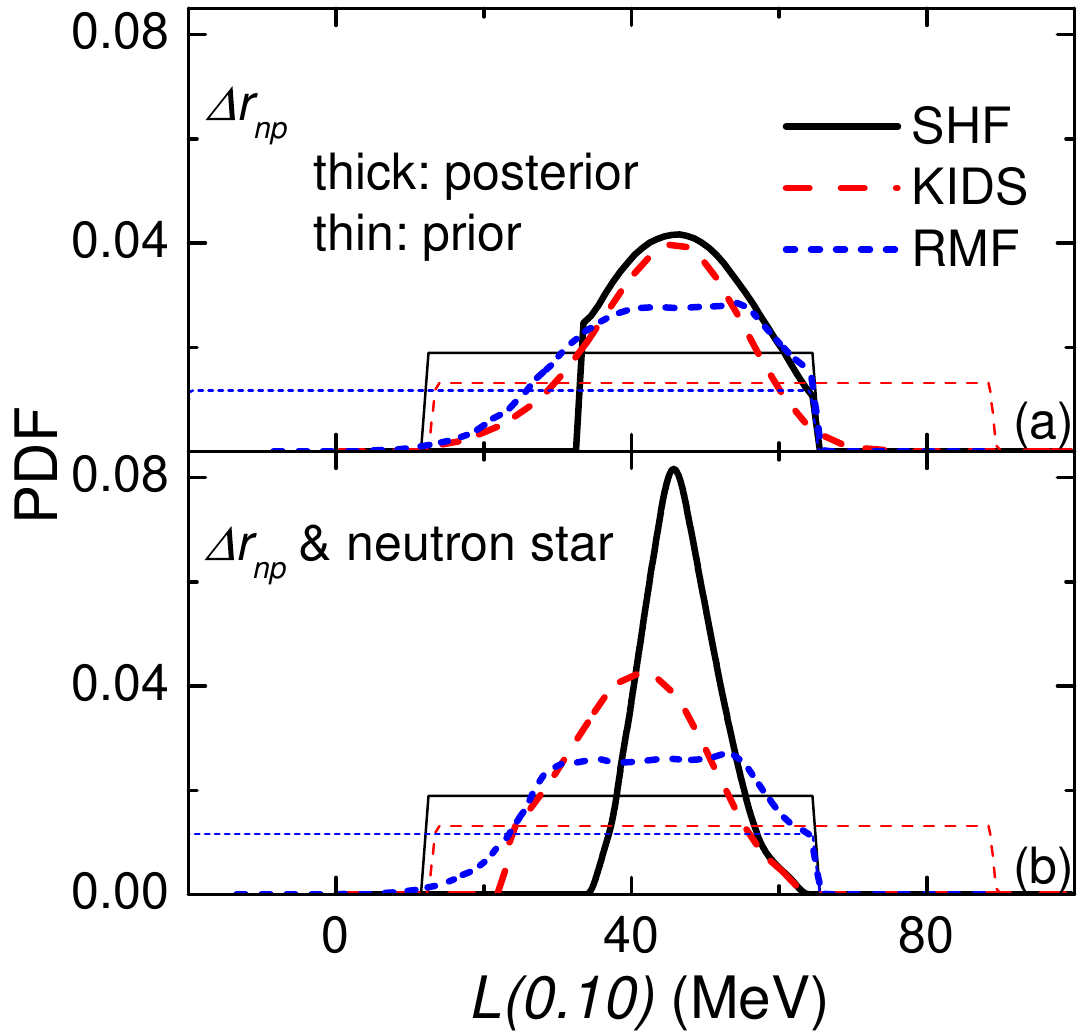}
\caption{\label{fig4} Posterior PDFs of the slope parameter at $\rho=0.10$ fm$^{-3}$ from the neutron-skin thickness data (a) and together with neutron-star observables (b), based on the standard SHF, KIDS, and RMF models.}
\end{figure}

\begin{figure}[!h]
\includegraphics[width=0.8\linewidth]{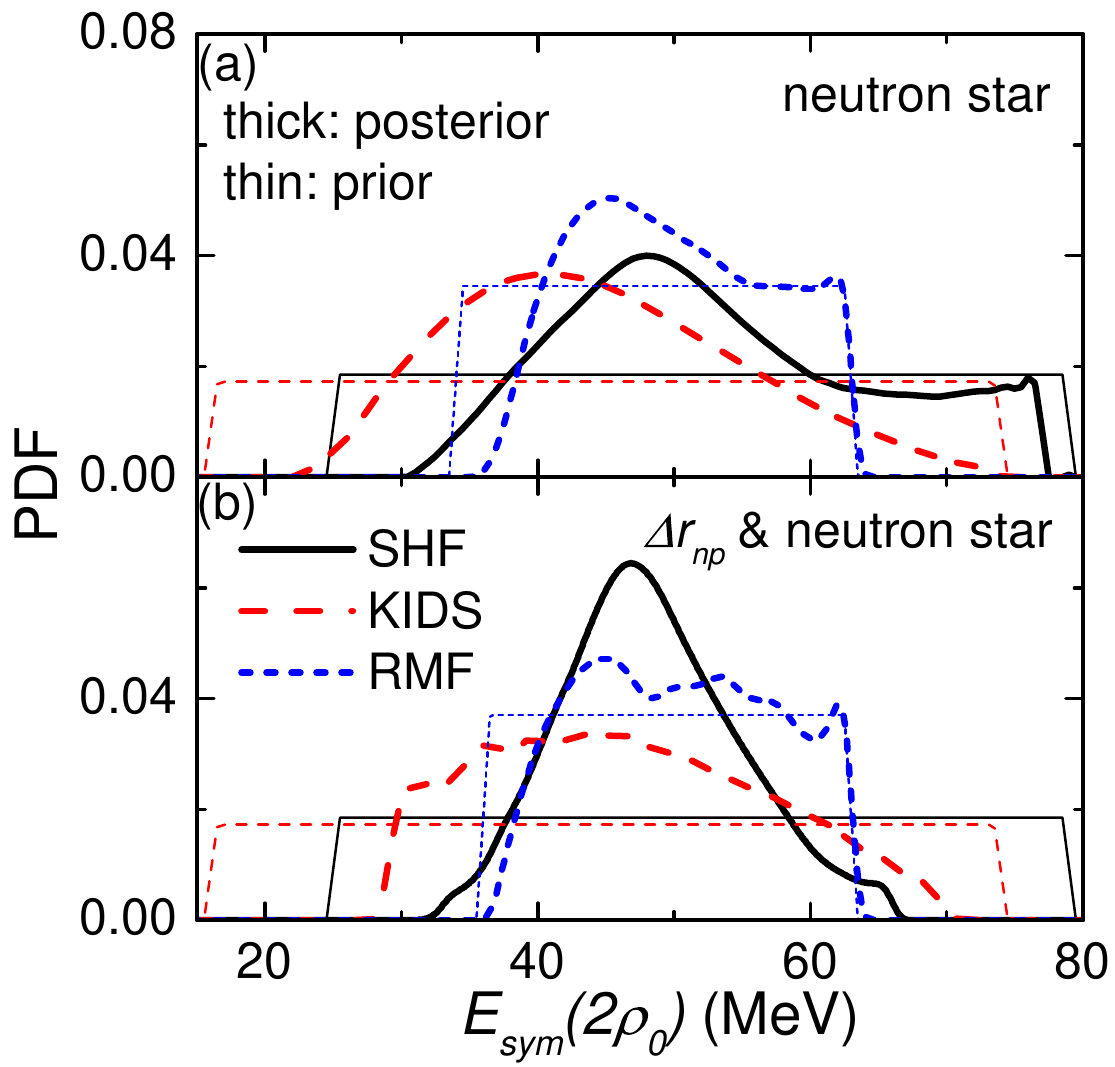}
\caption{\label{fig5} Posterior PDFs of the nuclear symmetry energy at $\rho=2\rho_0$ from neutron-star observables (a) and together with the neutron-skin thickness data (b), based on the standard SHF, KIDS, and RMF models.}
\end{figure}

It is seen from Fig.~\ref{fig3} that incorporating the additional constraint of neutron-star observables ($\Delta r_{np}$) may further constrain the symmetry energy at low (high) densities, compared to the constraints of $\Delta r_{np}$ (neutron-star observables) alone, according to the adopted energy-density functional. Here we illustrate the two situations separately. It has been shown that the $\Delta r_{np}$ data generally constrains the slope parameter $L(\rho^\star)=3\rho^\star (dE_{sym}/d\rho)_{\rho=\rho^\star}$ of the symmetry energy at $\rho^\star=0.10$ fm$^{-3}$~\cite{Xu:2020fdc,ZHANG2013234}, and the posterior PDFs of $L(0.10)$ based on three different models are compared in the upper panel of Fig.~\ref{fig4}. Model dependence is already observed here with only neutron-skin thickness data. If the constraint from neutron-star observables is further incorporated, the constraint on $L(0.10)$ is modified as shown in the lower panel of Fig.~\ref{fig4}, and the difference among the three models becomes even larger. For the standard SHF model, the PDF of $L(0.10)$ becomes sharper after incorporating the constraints from neutron-star observables. For the KIDS model, the peak of the PDF of $L(0.10)$ moves to a lower side, and this is consistent with the behavior of $L(0.16)$ shown in Fig.~\ref{fig2}(e). Actually, after including $K_{sym}$ as a independent EOS variable, the strong correlation between $\Delta r_{np}$ and $L(0.10)$ may not be rigorously valid (see, e.g., Ref.~\cite{PhysRevC.105.044305}). The behavior of the symmetry energy at suprasaturation densities is generally characterized by its value $E_{sym}(2\rho_0)$ at twice saturation density, with the fiducial value of about 47 MeV from various constraints as summarized in Ref.~\cite{Xie:2020tdo}. The posterior PDFs of $E_{sym}(2\rho_0)$ from neutron-star observables for the three models are compared in the upper panel of Fig.~\ref{fig5}. The too large values of $E_{sym}(2\rho_0)$ for the standard SHF model, which is also observed in Fig.~\ref{fig3}, is likely due to the constraint of the radii for large-mass neutron stars as well as the correlation between $L$ and $K_{sym}$, and they are ruled out after further incorporating the constraint from $\Delta r_{np}$ as shown in the lower panel of Fig.~\ref{fig5}. On the other hand, the PDFs of $E_{sym}(2\rho_0)$ for the KIDS and RMF models are not much affected after the constraint from $\Delta r_{np}$ is further incorporated. This is again due to the less-flexible feature of the standard SHF model compared to the KIDS and RMF models.

\section{Summary and outlook}
\label{sec:summary}

Based on the standard SHF, KIDS, and RMF models, we have studied the constraint on the density dependence of the symmetry energy $E_{sym}(\rho)$ from the neutron-skin thickness data of PREX and CREX as well as neutron-star data of GW170817, PSR J0030+0451, and PSR J0740+6620 using a Bayesian approach. Despite the soft and stiff symmetry energy favored respectively by the CREX and the PREX data, the Bayesian analysis is able to find a compromise for the $E_{sym}(\rho)$. While the neutron-skin thickness data (neutron-star observables) mostly constrain the $E_{sym}(\rho)$ at subsaturation (suprasaturation) densities, they more or less affect the constraint on the $E_{sym}(\rho)$ at suprasaturation (subsaturation) densities. For the RMF model, we found that the dependence of the neutron-skin thickness on the symmetry energy parameters can be quite different for a small Dirac effective mass ($m_s^\star/m<0.6$) compared to that for a large one ($m_s^\star/m>0.6$), as shown in Fig.~\ref{app}. While the key constraints on the $E_{sym}(\rho)$ from the present study can be found in Fig.~\ref{fig3}, the slope parameters at $\rho=0.10$ fm$^{-3}$ are constrained to be $47^{+4}_{-5}$ MeV for the standard SHF model, $41^{+9}_{-8}$ MeV for the KIDS model, and $43^{+13}_{-12}$ MeV for the RMF model, and the values of $E_{sym}(2\rho_0)$ are constrained to be $49^{+5}_{-7}$ MeV for the standard SHF model, $46^{+13}_{-9}$ MeV for the KIDS model, and $51^{+8}_{-7}$ MeV for the RMF model, within $68\%$ confidence intervals surrounding its mean value from both neutron-skin thickness data and neutron-star observables.

The constraints on the symmetry energy shown in the right column of Fig.~\ref{fig3} from the neutron-skin thickness data and the neutron-star observables are similar for the three adopted models. On the other hand, some model dependencies do exist, mainly due to the inclusion of higher-order EOS parameters and the difference between relativistic and non-relativistic models. While a model with a smaller number of free parameters is always favored, a more flexible model with more free parameters may be helpful in extracting detailed information of the nuclear interaction, as long as more constraints are incorporated from various observables based on the Bayesian analysis.

While the data from parity-violating electron-nucleus scattering experiments are less model-dependent, the large $1\sigma$ error bars for the PREX and CREX data reduce the constraining power on the slope parameter $L$ of the symmetry energy and the behavior of $E_{sym}(\rho)$. In future studies, we may adopt more nuclear structure data, including isotope binding energy difference and nucleus resonances, and then hopefully put a more stringent constraint on $E_{sym}(\rho)$.

\appendix

\section{Sensitivity investigation of neutron-skin thickness for the RMF model}
\label{appendix}

\begin{figure}[!h]
\includegraphics[width=1.0\linewidth]{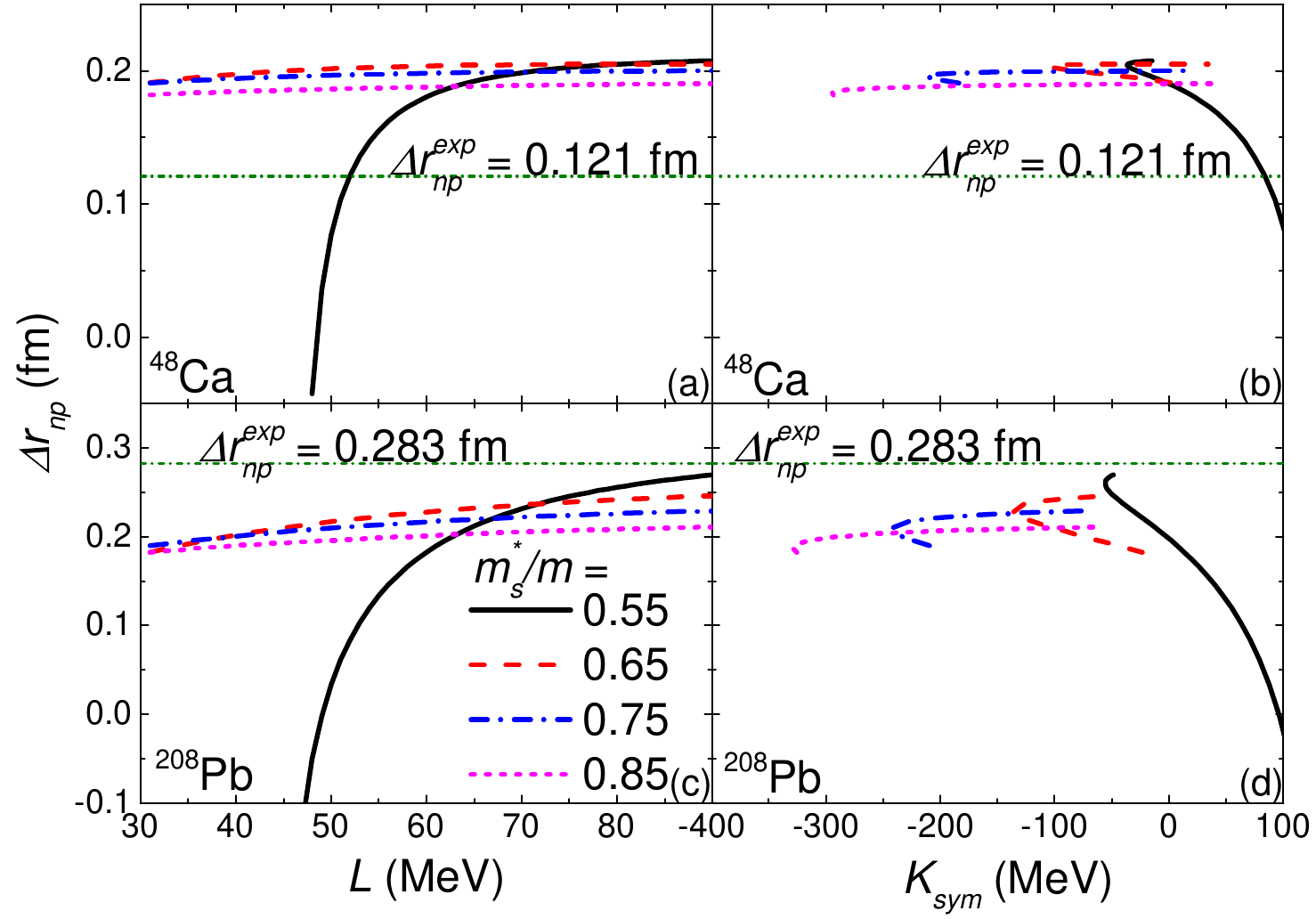}
\caption{\label{app} Dependence of the neutron-skin thickness in $^{48}$Ca [(a), (b)] and $^{208}$Pb [(c), (d)] on the slope parameter $L$ [(a), (c)] and the curvature parameter $K_{sym}$ [(b), (d)] of the symmetry energy for different values of isoscalar Dirac mass $m_s^\star$ in the RMF model.}
\end{figure}

To understand the abnormal posterior PDFs of the slope parameter $L$ and the curvature parameter $K_{sym}$ of the symmetry energy from the neutron-skin thickness $\Delta r_{np}$ data for the RMF model in Fig.~\ref{fig2}, we show in Fig.~\ref{app} the dependence of resulting $\Delta r_{np}$ in $^{48}$Ca and $^{208}$Pb on the corresponding symmetry energy parameters, for different values of isoscalar Dirac mass $m_s^\star$. The illustration is based an optimized parameter set ($K_0$, $Q_0$, $E_{sym}^0$, $L$, $m_s^\star$) that reproduces best the data of $^{48}$Ca and $^{208}$Pb, and then values of $L$ and $m_s^\star$ are varied. Since $K_{sym}$ is not chosen as an independent parameter in the RMF analysis, its value changes with $L$ for a fixed $m_s^\star$. For $m_s^\star/m=0.65$, 0.75, and 0.85, $\Delta r_{np}$ in both $^{48}$Ca and $^{208}$Pb increases almost linearly with increasing $L$, while their dependencies on $K_{sym}$ show non-monotonic behaviors. However, they are unable to reproduce the small $\Delta r_{np}^{exp}$ in $^{48}$Ca and the large $\Delta r_{np}^{exp}$ in $^{208}$Pb within the prior range of $L$. For $m_s^\star/m=0.55$, some abnormal behaviors are observed, and this is due to the too large $g_\rho^2/m_\rho^2$ value inversely obtained in the RMF model. It is seen that the $\Delta r_{np}$ in both $^{48}$Ca and $^{208}$Pb decreases dramatically with decreasing $L$ around $L=50-60$ MeV, or with increasing $K_{sym}$ at $K_{sym}>0$, for a small $m_s^\star/m$. At even smaller values of $L$, we are unable to get inversely the coefficients in the RMF model. For $^{48}$Ca, it is seen that the small $\Delta r_{np}$ can only be well reproduced by a small $m_s^\star/m$ and $L=50-60$ MeV, with the latter corresponding to the peak in the posterior PDF of $L$ in Fig.~\ref{fig1}(h). The range of $K_{sym}>0$ is also favored by the small $\Delta r_{np}^{exp}$ in $^{48}$Ca, corresponding to the posterior PDF of $K_{sym}$ in Fig.~\ref{fig1}(i). For $^{208}$Pb, even for a small $m_s^\star/m$, the large $\Delta r_{np}^{exp}$ favors a large $L$, so the posterior PDF of $L$ in Fig.~\ref{fig1}(h) looks normal. As can be seen from Fig.~\ref{app}(d), the large $\Delta r_{np}^{exp}$ in $^{208}$Pb favors a $K_{sym}$ around $-50$ MeV but disfavors a positive $K_{sym}$, consistent with the behavior in Fig.~\ref{fig1}(i).

\begin{acknowledgments}
We acknowledge helpful discussions with Panagiota Papakonstantinou. This work is supported by the Strategic Priority Research Program of the Chinese Academy of Sciences under Grant No. XDB34030000, the National Natural Science Foundation of China under Grant Nos. 12375125, 11922514, and 11475243, and the Fundamental Research Funds for the Central Universities.
\end{acknowledgments}

\bibliography{Nskin+Nstar_bayes}
\end{document}